\documentclass[12pt]{article}
\usepackage{amsfonts}
\usepackage{amsmath}
\usepackage{amssymb}
\usepackage{graphicx}
\usepackage{color}
\usepackage[all, knot]{xy}
\usepackage{tikz}

\usepackage{ulem}

\usepackage[utf8]{inputenc}
\usepackage{epstopdf}
\usepackage[footnotesize]{caption}
\usepackage{amsthm}
\usepackage{enumitem}
\usepackage{mathrsfs}

\usepackage[margin=3cm]{geometry}

\def \be {\begin{equation}}
\def \ee {\end{equation}}
\def \bea {\begin{eqnarray}}
\def \eea {\end{eqnarray}}
\def \nn {\nonumber}

\def \rr {\raise.35ex\hbox{\small $\prime$}\kern-.17em{\mbox{\large $\imath$}}}

\def \dels {\partial\kern-.6em /\kern.1em}
\def \As {{A\kern-.5em / \kern.5em}}
\def \Ds {D\kern-.7em / \kern.5em}

\def \ks {k\kern-.5em /}
\def \ls {l\kern-.5em /}

\newcommand{\ci}[1]{}

\newcommand{\ba}{\begin{eqnarray}}
\newcommand{\ea}{\end{eqnarray}}
\newcommand{\bal}{\begin{align}}
\newcommand{\eal}{\end{align}}
\newcommand{\bay}[1]{\left(\begin{array}{#1}}
\newcommand{\eay}{\end{array}\right)}

\newcommand{\hide}[1]{}

\newlist{axioms}{enumerate}{2}
\setlist[axioms,1]{label=\textbf{A\arabic{axiomsi}.}, ref=A\arabic{axiomsi}}
\setlist[axioms,2]{label=\textbf{A\arabic{axiomsi}\rlap{\myEnumCounter{axiomsii}}.},%
                   ref=A\arabic{axiomsi}\myEnumCounter{axiomsii},%
                   align=parleft,%
                   leftmargin=0em,%
                   itemsep=1.4ex,%
                   before={\stepcounter{axiomsi}}}

  \usetikzlibrary{decorations.markings}

\begin{document}

\begin{titlepage}
\begin{center}

\textbf{\LARGE
Cubic Action in Double Field Theory  
\vskip.3cm
}
\vskip .5in
{\large
Chen-Te Ma$^{a, b, c, d, e}$ \footnote{e-mail address: yefgst@gmail.com}
\\
\vskip 1mm
}
{\sl
$^a$
Guangdong Provincial Key Laboratory of Nuclear Science,\\ 
Institute of Quantum Matter,
South China Normal University, Guangzhou 510006, Guangdong, China.
\\
$^b$
School of Physics and Telecommunication Engineering,\\
 South China Normal University, 
 Guangzhou 510006, Guangdong, China.
 \\
 $^c$
Guangdong-Hong Kong Joint Laboratory of Quantum Matter,\\
 Southern Nuclear Science Computing Center, 
South China Normal University, Guangzhou 510006, Guangdong, China.
\\
$^d$
The Laboratory for Quantum Gravity and Strings,\\
 Department of Mathematics and Applied Mathematics,
University of Cape Town, Private Bag, Rondebosch 7700, South Africa.
\\
$^e$
Department of Physics and Center for Theoretical Sciences,\\
 National Taiwan University,
 Taipei 10617, Taiwan, R.O.C.
}
\\
\vskip 1mm
\vspace{30pt}
\end{center}
\newpage
\begin{abstract} 
We study target space theory on a torus for the states with $N_L+N_R=2$ through Double Field Theory. 
The spin-two Fierz-Pauli fields are not allowed when all spatial dimensions are non-compact. 
The massive states provide both non-vanishing momentum and winding numbers in the target space theory. 
To derive the cubic action, we provide the unique constraint for $N_L\neq N_R$ compatible with the integration by part.   
We first make a correspondence of massive and massless fields. 
The quadratic action is gauge invariant by introducing the mass term. 
We then proceed to the cubic order. 
The cubic action is also gauge invariant by introducing the coupling between the one-form field and other fields. 
The massive states do not follow the consistent truncation. 
One should expect the self-consistent theory by summing over infinite modes. 
Hence the naive expectation is wrong up to the cubic order.  
In the end, we show that the momentum and winding modes cannot both appear for only one compact doubled space.  
\end{abstract}
\end{titlepage}

\section{Introduction}
\label{sec:1}
\noindent 
String Theory describes the dynamics of a one-dimensional object. 
When a distance scale is much larger than the string scale, a string behaves like a point particle. 
The oscillation of a string generates the graviton and other fields. 
Because String Theory is ultraviolet (UV) complete, it is a candidate to provide a consistent Quantum Gravity. 
\\ 

\noindent 
String Theory compactified on a $d$-dimensional {\it torus} $T^d$ has a peculiar {\it O($d$, $d$; $\mathbb{Z}$)} symmetry \cite{Giveon:1991jj, Maharana:1992my}. 
Indeed, strings can propagate along and wrap non-contractible cycles in spacetime. 
It generates winding states that have no analogue for particle theories. 
Hence one can claim that O($d$, $d$; $\mathbb{Z}$) symmetry is {\it stringy}. 
For a rectangular torus, the momentum number $n_a$ is related to the momentum $p_a$ along the circle of radius $R_a$ by the following definition:
\bea
p_a\equiv\frac{n_a}{R_a}; \qquad n_{a}\in\mathbf{Z}; \qquad a=1, 2, \cdots, d; 
\eea
where $R_a$ is radius of a circle.  
The number of times the string winds around a circle is instead the winding number $m_a$, associated with the winding $\omega_a$ through: 
\bea
\omega_a\equiv m_a\frac{R_a}{\alpha^{\prime}}; \qquad m_{a}\in\mathbf{Z}, 
\eea 
where $\alpha^{\prime}$ is the Regge slope parameter. 
The characterization of spectrum in the non-compact Minkowski spacetime is by the following squared of mass:
\bea
 {\cal M}^{2} \equiv  - k^{2} = p^{2} + \omega^{2} +\frac{2}{\alpha^{\prime}}(N_L+N_R-2), 
 \label{m2lowdim}
\eea
where the momenta along the non-compact directions are $k^{\mu}$ while 
$N_L$ and $N_R$ are the numbers of left- and right-moving oscillators. 
We label the non-compact directions by $\mu=0, 1, \cdots, 25-d$.   
The spectrum is {\it invariant} under the target-space duality (T-duality) \cite{Buscher:1987sk, Buscher:1987qj}. 
Under the T-duality, exchanging $n_a$ and $m_a$ together with the exchange of $R_a$ and $\alpha^{\prime}/R_a$ \cite{Buscher:1987sk, Buscher:1987qj}. 
 Introducing simultaneously the coordinates $X^a$ for $T^d$ together with their duals $\tilde{X}_a$ can provide T-duality to a {\it manifest} symmetry.  
 This symmetric structure provides the manifest T-dual invariant formulation of the first-quantized String Theory \cite{Duff:1989tf, Tseytlin:1990nb, Siegel:1993xq, Siegel:1993th} and its target space \cite{Siegel:1993xq, Hohm:2010pp}. 
\\

\noindent
The closed string field theory (second-quantized String Theory) \cite{Zwiebach:1992ie, Hata:1986mz, Kugo:1992md, Alvarez:1996vt, Ghoshal:1991pu} on a {\it doubled} torus naturally exhibits a {\it manifest T-duality structure}. 
The manifest structure keeps the {\it momentum} and {\it winding} modes simultaneously. 
Therefore, it inspires {\it Double Field Theory} \cite{Hull:2009mi}. 
When restricting to {\it half} degrees of freedom, Double Field Theory reveals to be based on {\it Generalized Geometry} \cite{Gualtieri:2003dx, Hitchin:2004ut}. 
The tangent space at each point of the target space becomes a direct sum of the {\it tangent} and {\it cotangent} spaces. 
\\

\noindent
The on-shell string states have to satisfy the {\it level matching condition} 
\bea
L_0=\bar{L}_0
\eea
 and the free string on-shell condition. 
The $L_0$ and $\bar{L}_0$ are Virasoro operators. 
The free string on-shell condition determines the spectrum. 
When all the dimensions of the target space are uncompactified in the bosonic string theory, the level matching condition implies that
\bea
N_L=N_R, 
\eea   
where $N_L$ ($N_R$) is the number of left (right)-moving oscillators. 
\\

\noindent
In particular, massless states in the spectrum require $(N_L, N_R)=(1, 1)$. 
Therefore, the level matching condition gives a restriction. 
In the presence, instead, of compact dimensions in the target space, one obtains
\bea
N_L-N_R-\alpha^{\prime}p_a w^a=0. 
\eea 
With some {\it compact} directions, the momentum and winding appear in the level matching condition. 
Therefore, It leads to the {\it relaxation} of restriction for the choices of $N_L$ and $N_R$.
Double Field Theory rewrites the condition as in the following \cite{Hull:2009mi}: 
\bea
\partial^a\tilde{\partial}_af=-\frac{N_L-N_R}{\alpha^{\prime}}f\equiv-\frac{\lambda}{2}f,
\eea
where
\bea
\partial_a\equiv\frac{\partial}{\partial x^a}; \qquad 
\tilde{\partial}^a\equiv\frac{\partial}{\partial\tilde{x}_a}. 
\eea
One can choose $f$ to be the fluctuation fields or the gauge parameters. 
The derivative operator gives the momenta  
\bea
-i\partial_a\rightarrow\frac{n_a}{R_a}. 
\eea
The dual derivative operator gives the winding 
\bea
-i\tilde{\partial}^a\rightarrow \frac{m_aR_a}{\alpha^{\prime}}.
\eea 
Double Field Theory has necessarily to impose the constraint for getting the {\it gauge symmetry}.  
When one chooses $N_L=N_R$, the constraint is called {\it weak} constraint.  
When imposing such constraint to all products, it is called {\it strong} constraint.  
\\

\noindent
The {\it strong} constraint annihilates {\it half} of the degrees of freedom.  
When imposing this constraint, Double Field Theory {\it loses} the interaction between the {\it target space} and {\it dual space}. 
The weak constraint is more {\it relaxed} than the strong one. 
It is still unclear how to show gauge invariance with the weak constraint only. 
String Theory suggests that relaxing constraint is necessary on a space with compactified directions \cite{Betz:2014aia}. 
For spaces with one doubled compactified direction, the weak constraint is the same as the strong constraint \cite{Ma:2016vgq, Ma:2017jeq}.
\\

\noindent
Double Field Theory does not require the vanishing of ${\cal M}^2$ but, instead, of {\it another} definition of squared of mass: 
\bea
 M^{2} \equiv - (k^{2} + p^{2} + \omega^{2})=\frac{2}{\alpha^{\prime}}(N_L+N_R-2). \label{squaredm}
\eea 
Therefore, this theory keeps the $M^2=0$ states including all their momentum fluctuations and integrating everything else. 
Hence Double Field Theory {\it neglects} some {\it light} states from the {\it lower-dimensional} point of view. 
\\

\begin{figure}
\begin{center}
\includegraphics[width=1.\textwidth]{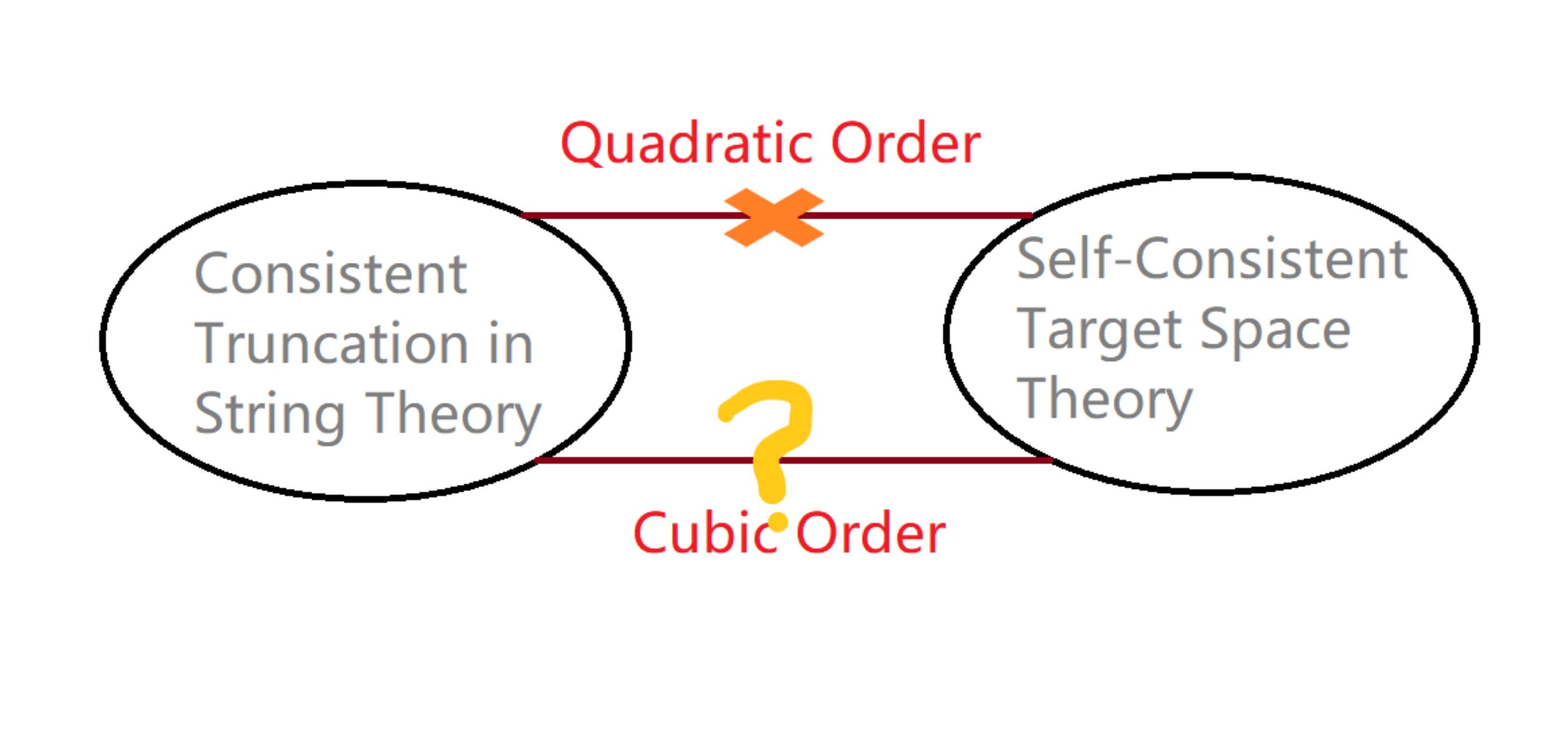}
\end{center}
\caption{We show the problem between String Theory and its target space.}
\label{Problem.pdf}
\end{figure}

\noindent
In this paper, we derive the cubic action from Double Field Theory for $N_L+N_R=2$. 
For $(N_L, N_R)=(1, 1)$, the fluctuation fields are: $h_{jk} (x^{\mu}, x^{a}, \tilde{x}_{a})$; 
$b_{jk} (x^{\mu}, x^{a}, \tilde{x}_{a})$; 
$\phi(x^{\mu}, x^{a}, \tilde{x}_{a})$, where $j, k=1, 2, \cdots, D$. 
Here $D$ is the number of compact and non-compact directions. 
In the bosonic string theory, it results $D=26$. 
For the massive states $(N_L, N_R)=(2, 0); (0, 2)$, the field strength associated with a one-form field replaces the $b_{jk}$. 
A well-known problem in Double Field Theory is no consistent truncation of massive states. 
The quadratic-order corresponds to the non-interacting theory. 
It is unnecessary to have the interaction between the different modes \cite{Ma:2019wbc}. 
Writing a self-consistent theory for each mode does not suffer any problems. 
The non-linear term should give the non-trivial constraints to the infinite summation for different numbers of $N_L$ and $N_R$. 
We show this problem in Fig. \ref{Problem.pdf}. 
We use the gauge symmetry with the constraint for each $\lambda$ to show the cubic action. 
Therefore, the consistent truncation does not bother with the self-consistency of target space theory. 
To summarize our results:
\begin{itemize}
\item We generalize the constraint for $\lambda\neq 0$ compatible with the integration by part. 
It shows the unique generalization. 
Products are commutative. 
For $\lambda=0$, the constraint annihilates half of the degrees of freedom. 
Massive states do not follow the strong constraint. 
\item The cubic action for $(N_L, N_R)=(1, 1)$ corresponds to $\lambda=0$. 
One can replace the massless two-form field with the field strength associated with the one-form field. 
Therefore, one can obtain the gauge transformation for the massive states by the one-to-one correspondence.  
Hence we can derive the cubic action for the spin-2 Fierz-Pauli fields or $(N_L, N_R)=(2, 0); (0, 2)$. 
We obtain the gauge-invariant action by introducing the additional coupling between the one-form field and other fields. 
\item We finally show that none of such new solutions allow both momentum and winding modes in the case of one double direction.  
\end{itemize}

\noindent
The organization of this paper is as follows: We generalize the constraint to $\lambda\neq 0$ in Sec.~\ref{sec:2}. 
We then show the closed gauge transformation in Sec.~\ref{sec:3}. 
The result of the quadratic and cubic orders for the action is in Sec.~\ref{sec:4} and Sec.~\ref{sec:5}. 
We show that one double direction is not enough for Double Field Theory in Sec.~\ref{sec:6}.   
In the end, we discuss our results and conclude in Sec.~\ref{sec:7}. 
We show the closed gauge transformation of the generalized metric in Appendix \ref{appa}. 
The derivation of the cubic action is in Appendix \ref{appb}. 
 
\section{Constraint}
\label{sec:2}
\noindent 
We define the constraint \cite{Hull:2009mi} and show the generalization to the multiple products.  
\\

\subsection{Definition} 
\noindent
The constraint is \cite{Hull:2009mi}
\bea
\partial_M\partial^M f=-\lambda f,
\eea
where 
\bea
\partial_M\equiv 
 \begin{pmatrix} \,\tilde{\partial}^m \, \\[0.6ex] {\partial_m } \end{pmatrix}.
\eea 
We label the doubled indices by $M=1, 2, \cdots, 2D$. 
We then raise or lower the doubled indices by the O($D$, $D$) metric \cite{Hull:2009mi, Gualtieri:2003dx, Hitchin:2004ut}
\bea
\eta= \begin{pmatrix} 0& I \\ I& 0 \end{pmatrix}.
\eea 
We implement the constraint on the momentum space as follows. 
The star product provides the convenient notation
\bea
A*1\equiv\sum_K A_K\exp(iKX)\delta_{KK,\lambda}, 
\eea
where 
\bea
X^M\equiv (\tilde{x}_j, x^j), \qquad K^M\equiv (p_j, \omega^j);
\eea
\bea
KK\equiv K^MK_M, \qquad KX\equiv K^MX_M. 
\eea 
This notation distinguishes $A$ from $A*1$. 
Although the notation is unique, it is not manageable for being explicitly written all the time.  
Therefore, we will assume that all doubled fields satisfy the constraints in deriving quadratic and cubic actions. 
The $x^j$'s and $\tilde{x}_j$'s are the compact coordinate and the dual, respectively. 
The $p_j$'s and $\omega^j$'s are the momenta and windings along with the compact directions, respectively. 
This product imposes a constraint on $A$
\bea
\partial^M\partial_M(A*1)=-\lambda (A*1)
\eea 
that is the only one required in deriving the quadratic action. 
A higher number of products is necessary when interested in higher orders. 
Therefore, we first discuss the quadratic products for the quadratic action. 

\subsection{Quadratic Action}
\noindent
The integration by part shows the constraint
\bea
K_A+K_B=0
\eea 
for the quadratic term 
\bea
\int[dxd\tilde{x}]\ A*B. 
\eea 
Therefore, we obtain 
\bea
K_AK_B=-K_A^2=-\lambda.
\eea
Hence the quadratic products are:
\bea
A*B=B*A\equiv\sum_{K_A, K_B}A_{K_A}B_{K_B}\exp\big(i(K_A+K_B)X\big)\delta_{K_AK_A, \lambda}\delta_{K_BK_B, \lambda}\delta_{K_AK_B, -\lambda}.
\eea 
When $\lambda$ vanishes, the constraint reduces to the strong constraint. 
The quadratic products satisfy the constraint
\bea
\partial_M\partial^M(A*B)=0
\eea
for each $\lambda$. 
The restriction implies that the quadratic action satisfies the constraint as the strong constraint case. 
Because the field still satisfies the relaxed constraint, the constraint of quadratic products does not prohibit the non-orthogonality of momenta.
Integrating by part also provides the following equality
\bea
\int[dxd\tilde{x}]\ A*B=\int[dxd\tilde{x}]\ AB. 
\eea 
For massive states, $\lambda$ still leads to the possibility of going beyond String Theory. 

\subsection{Cubic Action}
\noindent
For the cubic action, the integration by part also shows the similar constraint 
\bea
K_A+K_B+K_C=0 
\eea
from the following integration:
\bea
\int[dxd\tilde{x}]\ (A*B)*C; \qquad \int[dxd\tilde{x}]\ A*(B*C). 
\eea 
Each momentum satisfies the condition:
\bea
K_A^2=K_B^2=K_C^2=\lambda. 
\eea
Hence we obtain: 
\bea
&&
K_AK_B=K_AK_C=K_BK_C=-\frac{\lambda}{2},
\nn\\
&&
(K_A+K_B)K_C=(K_A+K_C)K_B=(K_B+K_C)K_A=-\lambda; 
\eea
\bea
\partial_MA*\partial^MB=\frac{\lambda}{2}A*B, \qquad \partial_MA*\partial^MC=\frac{\lambda}{2}A*C, \qquad 
\partial_MB*\partial^MC=\frac{\lambda}{2}B*C,
\eea
\bea
&&
\partial_M(A*B)*\partial^MC=\lambda(A*B)*C, 
\nn\\
&&
 \partial_M(A*C)*\partial^MB=\lambda(A*C)*B, 
 \nn\\
 &&
\partial_M(B*C)*\partial^MA=\lambda(B*C)*A. 
\eea 
The triple products are associative: 
\bea
&&A*(B*C)
\nn\\
&=&(A*B)*C
\nn\\
&\equiv&\sum_{K_A, K_B, K_C}A_{K_A}B_{K_B}C_{K_C}
\exp\big(i(K_A+K_B+K_C)X\big)
\nn\\
&&\times\delta_{K_AK_A, \lambda}\delta_{K_BK_B, \lambda}\delta_{K_CK_C, \lambda}
\delta_{K_AK_B, -\frac{\lambda}{2}}\delta_{K_AK_C, -\frac{\lambda}{2}}.
\eea 
The triple products also satisfy the constraint:
\bea
\partial^M\partial_M\big(A*(B*C)\big)=0
\eea
for each $\lambda$. 
It implies that the cubic action has a similar situation to the quadratic action.
One can unify the cubic and quadratic action by writing
\bea
\partial_M(A*1)\partial^MC=\lambda(A*1)*C.
\eea
For the convenient notation, we will not write the quadratic action in terms of the triple products. 

\subsection{Quadruple Products}
\noindent 
For the quadruple products, the integration by part shows: 
\bea
K_A+K_B+K_C+K_D=0,
\eea
and the level matching condition gives:
\bea
K_A^2=K_B^2=K_C^2=K_D^2=\lambda. 
\eea 
The momenta satisfy: 
\bea
&&
K_AK_B+K_AK_C+K_BK_C=K_AK_B+K_AK_D+K_BK_D
\nn\\
&=&K_AK_C+K_AK_D+K_CK_D=K_BK_C+K_BK_D+K_CK_D
\nn\\
&=&-\lambda.
\eea
It is easy to observe that: 
\bea
(K_A+K_B)K_C=-\lambda-K_AK_B; \qquad 
(K_B+K_C)K_A=-\lambda-K_BK_C.
\eea
Therefore, the triple products are not associated here with the quadruple action. 
A general simplification is only up to our concern (cubic action).

\section{Gauge Transformation}
\label{sec:3}
\noindent
We apply the constraint to realize the non-linear gauge transformation. 
The product is only commutative up to triple products. 
It implies that some field configurations should have the gauge symmetry, defined by the commutative product. 
Hence the non-linear realization should provide non-trivial evidence to the constraint. 
The non-vanishing $\lambda$ goes beyond the strong constraint. 
Showing a closed gauge transformation is already non-trivial. 
It is also general enough for a cubic action.   
\\

\noindent 
Here we discuss the generalized metric \cite{Duff:1989tf} and the scalar dilaton: 
\bea
{\cal H}_{MN}\equiv
\begin{pmatrix}
g^{-1}& -g^{-1}*{\mbox b}
\\
{\mbox b}*g^{-1}& g-{\mbox b}*g^{-1}*{\mbox b} \label{HMN}  
\end{pmatrix}; \qquad 
\Phi = e^{-2 \phi} * \sqrt{| \mbox{det} g |}.
\eea 
Here $g_{jk}$ is a symmetric tensor field, and ${\mbox b}_{jk}$ is an anti-symmetric tensor field. 
The generalized metric combines the symmetric and anti-symmetric components into one symmetric structure.
\\

\noindent 
The gauge transformations are:
\bea
\delta_{\xi}{\cal H}^{MN}
&=&\xi^{P}*\partial_{P}{\cal H}^{MN}+\big(\partial^{M}\xi_{P}-\partial_{P}\xi^{M}\big)*{\cal H}^{PN}+\big(\partial^{N}\xi_{P}-\partial_{P}\xi^{N})*{\cal H}^{MP};
\nn\\
\delta_{\xi} \Phi
&=&-\frac{1}{2}\partial_{M}\xi^{M}+\xi^{M}*\partial_{M}\Phi,     \label{GT}
\eea
where the gauge parameter $\xi_{P}$ is 
\bea
\xi^P\equiv\begin{pmatrix}
\tilde{\xi}_j 
\\
\xi^j
\end{pmatrix}.
\eea 
The vector field $\xi^j$ generates the diffeomorphism while another component provides a dual one. 
The scalar dilaton for the fluctuation level becomes 
\bea
\exp(-2\Phi)\equiv\exp(-2\phi)\sqrt{-\det h_{jk}}.
\eea 
\\

\noindent 
We show the closed gauge transformation for the scalar dilaton. 
The computation is as in the following: 
\bea
&&\lbrack\delta_{\xi_1}, \delta_{\xi_2}\rbrack_* \Phi
\nn\\
&\equiv&-\frac{1}{2}\xi_2^M*\partial_M\partial_P\xi_1^P
+\xi_2^M*\partial_M\xi_1^P*\partial_P\Phi
+\frac{1}{2}\xi_1^M*\partial_M\partial_P\xi_2^P
-\xi_1^M*\partial_M\xi_2^P*\partial_P\Phi
\nn\\
&&
+\xi_2^M*\xi_1^P*\partial_M\partial_P\Phi
-\xi_1^M*\xi_2^P*\partial_M\partial_P\Phi
\nn\\
&=&-\frac{1}{2}\xi_2^M*\partial_M\partial_P\xi_1^P
+\xi_2^M*\partial_M\xi_1^P*\partial_P\Phi
+\frac{1}{2}\xi_1^M*\partial_M\partial_P\xi_2^P
-\xi_1^M*\partial_M\xi_2^P*\partial_P\Phi
\nn\\
&=&\frac{1}{2}\partial_M\lbrack\xi_1, \xi_2\rbrack_C^M-\lbrack\xi_1, \xi_2\rbrack^M_C\partial_M\Phi
\nn\\
&=&-\delta_{\lbrack\xi_1, \xi_2\rbrack_C}\Phi.
\eea
Here the $C$-bracket is: 
\bea
\lbrack \xi_{1}, \xi_{2} \rbrack_{C}^{M} \equiv  \xi^{N}_{\lbrack 1}*\partial_{N} \xi^{M}_{2\rbrack} 
- \frac{1}{2}  \xi^{P}_{\lbrack 1 }*\partial^{M} \xi_{2 \rbrack P}, \ 
\lbrack j, k\rbrack \equiv jk -kj. 
\eea 
We used following results in the calculation:
\bea
&&\partial_M\lbrack\xi_1, \xi_2\rbrack_C^M
\nn\\
&=&\partial_M\bigg(\xi_1^N*\partial_N\xi_2^M-\xi_2^N*\partial_N\xi_1^M-\frac{1}{2}\eta^{MN}\eta_{PQ}\xi_1^P*\partial_N\xi_2^Q+\frac{1}{2}\eta^{MN}\eta_{PQ}\xi_2^P*\partial_N\xi_1^Q\bigg)
\nn\\
&=&\partial_M\xi_1^N*\partial_N\xi_2^M+\xi_1^N*\partial_M\partial_N\xi_2^M
-\partial_M\xi_2^N*\partial_N\xi_1^M-\xi_2^N*\partial_M\partial_N\xi_1^M
\nn\\
&&-\frac{1}{2}\partial_M\xi_1^P*\partial^M\xi_{2,P}-\frac{1}{2}\xi_1^P*\partial_M\partial^M\xi_{2,P}
+\frac{1}{2}\partial_M\xi_2^P*\partial^M\xi_{1,P}+\frac{1}{2}\xi_2^P*\partial_M\partial^M\xi_{1,P}
\nn\\
&=&\xi_1^N*\partial_M\partial_N\xi_2^M-\xi_2^N*\partial_M\partial_N\xi_1^M
-\frac{\lambda}{4}\xi_1^P*\xi_{2,P}+\frac{\lambda}{2}\xi_1^P*\xi_{2, P}+\frac{\lambda}{4}\xi_2^P*\xi_{1, P}-\frac{\lambda}{2}\xi_2^P*\xi_{1,P}
\nn\\
&=&
\xi_1^N*\partial_M\partial_N\xi_2^M-\xi_2^N*\partial_M\partial_N\xi_1^M;
\eea
\bea
&&-\lbrack\xi_1, \xi_2\rbrack^M_C*\partial_M\Phi
\nn\\
&=&-\bigg(\xi_1^N*\partial_N\xi_2^M-\xi_2^N*\partial_N\xi_1^M-\frac{1}{2}\eta^{MN}\eta_{PQ}\xi_1^P*\partial_N\xi_2^Q
+\frac{1}{2}\eta^{MN}\eta_{PQ}\xi_2^P*\partial_N\xi_1^Q\bigg)*\partial_M\Phi
\nn\\
&=&\xi_2^M*\partial_M\xi_1^P*\partial_P\Phi-
\xi_1^M*\partial_M\xi_2^P*\partial_P\Phi
\nn\\
&&
+\frac{1}{2}\xi_1^P*\partial^M\xi_{2, P}*\partial_M\Phi
-\frac{1}{2}\xi_2^P*\partial^M\xi_{1, P}*\partial_M\Phi
\nn\\
&=&\xi_2^M*\partial_M\xi_1^P*\partial_P\Phi
-\xi_1^M*\partial_M\xi_2^P*\partial_P\Phi.
\eea
The quadratic products have a non-unique notation in our paper. 
Here we use:  
\bea
\partial_M{\cal A}*\partial^M{\cal C}=\frac{\lambda}{2}{\cal A}*{\cal C}\neq\lambda{\cal A}*{\cal C}.
\eea
For the generalized metric, the gauge transformation is also closed similarly
\bea
\lbrack\delta_{\xi_1}, \delta_{\xi_2}\rbrack_*{\cal H}^{MN}
=-\delta_{\lbrack\xi_1, \xi_2\rbrack_*}{\cal H}^{MN}. 
\eea
We write the details in Appendix \ref{appa}. 
Our result implies that the gauge transformations of the cubic action are also closed under the constraint. 

\section{Quadratic Theory} 
\label{sec:4}
\noindent 
We first discuss the field content for $N_L+N_R=2$. 
We then derive the quadratic action from the gauge transformation. 
Both appearances of momentum and winding modes generate a mass term.

\subsection{Field Content} 
\noindent
The choices of $N_L$ and $N_R$ are $(N_L, N_R)=(1, 1); (2, 0); (0, 2)$. 
For the massless state $(N_L, N_R)=(1, 1)$, the first-excited state is $\alpha_{-1}^{j}\bar{\alpha}_{-1}^{k}|0\rangle$, where $\alpha_{-1}^{j}$ is the left creation operator, and $\bar{\alpha}_{-1}^{k}$ is the right one. 
It generates a symmetric traceless tensor field $h_{jk}$ with $(D-2)(D-1)/2-1$ physical degrees of freedom; 
a scalar field $\phi$ with one physical degree of freedom; 
an anti-symmetric tensor field $b_{jk}$ with $(D-2)(D-3)/2$ physical degrees of freedom. 
The number of total physical degrees of freedom is $(D-2)^2$. 
It forms the massless particle's representation of SO($D-2$). 
The spin-2 Fierz-Pauli fields with the Lorentz symmetry form the representation of SO($D-1$). 
Later we will show that the massive states $(N_L, N_R)=(2, 0); (0,2)$ correspond to the spin-2 Fierz-Pauli fields. 
\\

\noindent
For $(N_L, N_R)=(2, 0)$, the first excited states are $\alpha_{-1}^j\alpha_{-1}^k|0\rangle$ and $\alpha_{-2}^j|0\rangle$. 
The operator $\alpha_{-1}^j\alpha_{-1}^k$ generates a symmetric traceless tensor field $h_{jk}$ with $(D-2)(D-1)/2-1$ physical degrees of freedom; 
a scalar field $\phi$ with one physical degree of freedom. 
The operator $\alpha_{-2}^j$ generates a one-form gauge field $A_j$ with $D-2$ physical degrees of freedom. 
Hence the number of total physical degrees of freedom is $(D-2)(D+1)/2$. 
This number is the same as the dimension of the traceless symmetric tensor of SO($D-1$). 
We conclude that the massive state is spin-2. 
For $(N_L=0, N_R=2)$, we replace the left creation field with the right one. 
The result remains the same. 
We determine the allowed momentum and winding numbers by the level matching condition
\bea
N_L-N_R=n_am_a. 
\eea 
For massive states with one compactified direction, the allowed momentum and winding numbers are:
\bea
&&
(N_L, N_R)=(2, 0), \ (n, m)=(1, 2), (2, 1), (-1, -2), (-2, -1); 
\nn\\
&&
(N_L, N_R)=(0, 2), \ (n, m)=(-1, -2), (-2, -1), (1, -2), (2, -1).
\eea

\subsection{Quadratic Action}
\noindent
We consider the fluctuations around the constant symmetric and the vanishing anti-symmetric backgrounds. 
The quadratic action for the massless state $(N_L=1, N_R=1)$ is \cite{Hull:2009mi}
\bea
&&S^{(2)}_{DFT}
\nn\\
&=&\frac{1}{16\pi G_{N}}\int [dxd\tilde{x}] \ \bigg(\frac{1}{4}h^{jk}\partial_{l}\partial^{l}h_{jk}
+\frac{1}{2}\partial^{j}h_{jk}\partial_{l}h^{lk}
-2 \Phi \partial^{j}\partial^{k}h_{jk}-4 \Phi \partial^{j}\partial_{j} \Phi
\nn\\
&&+\frac{1}{4}h^{jk}\tilde{\partial}_{l}\tilde{\partial}^{k}h_{jk}
+\frac{1}{2}\tilde{\partial}^{j}h_{jl}\tilde{\partial}_{k}h^{k l}
+2 \Phi  \tilde{\partial}^{j}\tilde{\partial}^{k}h_{jk}-4 \Phi \tilde{\partial}^{j}\tilde{\partial}_{j} \Phi
\nn\\
&&+\frac{1}{4}b^{jk}\partial^{l}\partial_{l}b_{jk}
+\frac{1}{2}\partial^{k}b_{jk}\partial_{l}b^{j l}
+\frac{1}{4}b^{jk}\tilde{\partial}^{l}\tilde{\partial}_{l}b_{jk}
+\frac{1}{2}\tilde{\partial}^{k}b_{jk}\tilde{\partial}_{l}b^{jl}
\nn\\
&&+(\partial_{k}h^{j k})\tilde{\partial}^{l}b_{jl}
+(\tilde{\partial}^{l}h_{jl  })\partial_{k}b^{jk}
-4 \Phi \partial^{j}\tilde{\partial}^{k}b_{jk}\bigg),   \label{s11}
\eea
where $G_{N}$ is the gravitational constant. 
The integration $\int [dxd\tilde{x}]$ is over all of the $n+2d$ coordinates of $\mathbb{R}^{n-1,1} \times T^{2d}$. 
The periodic coordinates of the doubled torus are $(x^{a}, \tilde{x}_{a})$. 
We will write an analogous action for the massive states from $S^{(2)}_{DFT}$. 
Here all fluctuation fields also satisfy the constraint. 
Therefore, we should use the star product in the quadratic action. 
We should also use the triple way to rewrite the quadratic product for a unique notation. 
However, keeping in mind that the double fields always satisfy the constraint of the cubic action.  
We will use the simplified notation (but non-unique) to simplify the derivation.  
\\

\noindent 
The gauge transformation of $S^{(2)}_{DFT}$ is a linear doubled diffeomorphism \cite{Hull:2009mi}: 
\bea
\delta h_{jk} & = & \partial_{j}\epsilon_{k}+\partial_{k}\epsilon_{j}
+\tilde{\partial}_{j}\tilde{\epsilon}_{k}+\tilde{\partial}_{k}\tilde{\epsilon}_{j}; 
\nn\\
\delta b_{jk} & = & -\big(\tilde{\partial}_{j}\epsilon_{k}-\tilde{\partial}_{k}\epsilon_{j}\big) 
-\big(\partial_{j}\tilde{\epsilon}_{k}-\partial_{k}\tilde{\epsilon}_{j}\big);  
\nn\\
\delta \Phi & = & -\frac{1}{2} ( \partial_{j}\epsilon^{j}-\tilde{\partial}_{j}\tilde{\epsilon}^{j}). 
\label{gaugetransf}
\eea 
When the gauge parameters and fields are independent of $\tilde{x}_j$, the doubled diffeomorphism reduces to the standard one. 
\\
   
\noindent
The fields associated with $(N_L=2, N_R=0)$ and $(N_L=0, N_R=2)$ are: 
a symmetric traceless tensor $h_{jk}(x^{\mu}, x^{a}, \tilde{x}_{a})$; 
a one-form gauge field $A_j(x^{\mu}, x^{a}, \tilde{x}_{a})$; 
a scalar dilaton field $\Phi (x^{\mu}, x^{a}, \tilde{x}_{a})$. 
We define the field strength associated with the one-form gauge field $A_j$ is: 
\bea
b_{jk} =  -\big(\tilde{\partial}_{j}A_{k}-\tilde{\partial}_{k}A_{j}\big)
+\big(\partial_{k}A_{j}-\partial_{j}A_{k}\big).   \label{bjk}
\eea 
The gauge transformation of $A_{j}$ is:
\bea
\delta A_{j}  =   \epsilon_{j} = \tilde{\epsilon}_{j}.
\eea 
Therefore, the field strength has the same gauge transformation as the anti-symmetric field for $(N_L, N_R)=(1, 1)$. 
This result implies $\epsilon_{j} = \tilde{\epsilon}_{j}$. 
Later we will show that the gauge invariance leads to the assumption.  
Hence a one-to-one correspondence of the fields for the massive and massless states \cite{Hull:2009mi}. 
It allows us to consider the quadratic action \eqref{s11} as our starting point for the massive states. 
When one applies the correspondence to the massless theory, we also change the notation
\bea
S^{(2)}_{DFT} \rightarrow \tilde{S}^{(2)}_{new}
\eea
Here we distinguish the massless and massive quadratic actions because their field contents are different.  
\\

\noindent 
Since $\tilde{S}^{(2)}_{new}$ is gauge invariant for $\lambda=0$, it is only necessary to show the calculation for $\lambda\neq 0$. 
The gauge transformation for the $\tilde{S}^{(2)}_{new}$ is: 
\bea
&&\delta \tilde{S}^{(2)}_{new}
\nn\\
&=&\frac{1}{16\pi G_N}\int [dxd\tilde{x}]\ \bigg(\partial^{j}\tilde{\partial}_{j}\tilde{\epsilon}_{l}\partial_{k}h^{kl}
-4\Phi\partial^{j}\partial^{k}\tilde{\partial}_{j}\tilde{\epsilon}_{k}
+
\partial^{j}\tilde{\partial}_{j}\epsilon_{l}\tilde{\partial}_{k}h^{kl}
+4\Phi\tilde{\partial}^{j}\tilde{\partial}^{k}\partial_{j}\epsilon_{k}
\nn\\
&&+\partial^{k}\tilde{\partial}_{k}\epsilon_{j}\partial_{l}b^{jl}
+\partial^{k}\tilde{\partial}_{k}\tilde{\epsilon}_{j}\tilde{\partial}_{l}b^{jl}
+(\partial_{l}\tilde{\partial}^{l}\tilde{\epsilon}^{j})\tilde{\partial}^{k}b_{jk}
+(\partial_{l}\tilde{\partial}^{l}\epsilon^{j})\partial^{k}b_{jk}
\nn\\
&&
+(\partial_{l}h^{jl})\tilde{\partial}^{k}\partial_{k}\tilde{\epsilon}_{j}
+(\tilde{\partial}_{l}h^{jl})\tilde{\partial}^{k}\partial_{k}\epsilon_{j}
+4\Phi\partial^{j}\tilde{\partial}^{k}\tilde{\partial}_{j}\epsilon_{k}
-4\Phi\partial^{j}\tilde{\partial}^{k}\partial_{k}\tilde{\epsilon}_{j}\bigg)
\nn\\
&=&\frac{1}{16\pi G_N}\int [dxd\tilde{x}]\ \bigg(-\frac{\lambda}{2}\tilde{\epsilon}_{l}\partial_{k}h^{kl}+2\lambda \Phi \partial^{k}\tilde{\epsilon}_{k}-\frac{\lambda}{2}\epsilon_{l}\tilde{\partial}_{k}h^{kl}-2\lambda \Phi\tilde{\partial}^{k}\epsilon_{k}
\nn\\
&&-\frac{\lambda}{2}\epsilon_{j}\partial_{l}b^{jl}-\frac{\lambda}{2}\tilde{\epsilon}_{j}\tilde{\partial}_{l}b^{jl}-\frac{\lambda}{2}\tilde{\epsilon}^{j}\tilde{\partial}^{k}b_{jk}
-\frac{\lambda}{2}\epsilon^{j}\partial^{k}b_{jk}
\nn\\
&&-\frac{\lambda}{2}\tilde{\epsilon}_{j}\partial_{l}h^{jl}
-\frac{\lambda}{2}\epsilon_{j}\tilde{\partial}_{l}h^{jl}-2\lambda \Phi\tilde{\partial}^{k}\epsilon_{k}+2\lambda \Phi\partial^{j}\tilde{\epsilon}_{j}\bigg)
\nn\\
&=&\frac{1}{16\pi G_N}\int [dxd\tilde{x}]\ \bigg(-\lambda\tilde{\epsilon}_{l}\partial_{k}h^{kl}-\lambda\epsilon_{l}\tilde{\partial}_{k}h^{kl}
-\lambda\epsilon_{j}\partial_{l}b^{jl}-\lambda\tilde{\epsilon}_{j}\tilde{\partial}_{l}b^{jl}
\nn\\
&&+4\lambda \Phi\partial^{k}\tilde{\epsilon}_{k}-4\lambda \Phi\tilde{\partial}^{k}\epsilon_{k}
\bigg)
\nn\\
&=&\frac{1}{16\pi G_N}\int [dxd\tilde{x}]\ \bigg(-\lambda\tilde{\epsilon}_{j}\partial_{k}(h^{jk}+b^{jk})
-\lambda\epsilon_{j}\tilde{\partial}_{k}(h^{jk}+b^{jk})+4\lambda \Phi(\partial^{k}\tilde{\epsilon}_{k}-\tilde{\partial}^{k}\epsilon_{k})\bigg)
\nn\\
&=&\frac{1}{16\pi G_N}\int [dxd\tilde{x}]\ \bigg(\lambda (h^{jk}+b^{jk})(\partial_{k}\tilde{\epsilon}_{j}+\tilde{\partial}_{k}\epsilon_{j})+4\lambda \Phi(\partial^{k}\tilde{\epsilon}_{k}-\tilde{\partial}^{k}\epsilon_{k})\bigg)
\nn\\
&=&\frac{1}{16\pi G_N}\int [dxd\tilde{x}]\ \bigg(\frac{\lambda}{2} b^{jk}(\partial_{k}\tilde{\epsilon}_{j}-\partial_{j}\tilde{\epsilon}_{k}+\tilde{\partial}_{k}\epsilon_{j}
-\tilde{\partial}_{j}\epsilon_{k})
+\lambda h^{jk}(\partial_{k}\tilde{\epsilon}_{j}+\tilde{\partial}_{k}\epsilon_{j})
\nn\\
&&+4\lambda \Phi(\partial^{k}\tilde{\epsilon}_{k}-\tilde{\partial}^{k}\epsilon_{k})\bigg)
\nn\\
&=&\frac{1}{16\pi G_N}\int [dxd\tilde{x}]\ \bigg(\frac{\lambda}{4}\delta(b^{jk}b_{jk})+\lambda h^{jk}(\partial_{k}\tilde{\epsilon}_{j}+\tilde{\partial}_{k}\epsilon_{j})+4\lambda \Phi(\partial^{k}\tilde{\epsilon}_{k}-\tilde{\partial}^{k}\epsilon_{k})\bigg).
\eea 
The result of gauge transformation shows the non-invariant terms. 
The constraint
\bea
\partial^j\tilde{\partial}_jf=-\frac{\lambda}{2}f
\eea
cannot annihilate these terms.
Hence it is necessary to introduce the new term to restore the gauge symmetry. 
Now we assume that 
\bea
\epsilon_{j}=\tilde{\epsilon}_{j}.
\eea
We then obtain:
\bea
&&\delta \tilde{S}^{(2)}_{new}
\nn\\
&=&\frac{1}{16\pi G_N}\int [dxd\tilde{x}]\ \bigg(\frac{\lambda}{4}\delta(b^{jk}b_{jk})+\lambda h^{jk}(\partial_{k}\tilde{\epsilon}_{j}+\tilde{\partial}_{k}\epsilon_{j})+4\lambda \Phi(\partial^{j}\tilde{\epsilon}_{j}-\tilde{\partial}^{j}\epsilon_{j})\bigg)
\nn\\
&=&\frac{1}{16\pi G_N}\int [dxd\tilde{x}]\ \bigg(\frac{\lambda}{4}\delta(b^{jk}b_{jk})
+\frac{\lambda}{2} h^{jk}\delta h_{jk}-8\lambda \Phi\delta \Phi\bigg)
\nn\\
&=&\frac{1}{16\pi G_N}\int [dxd\tilde{x}]\ \bigg(\frac{\lambda}{4}\delta(b^{jk}b_{jk})+\frac{\lambda}{4}\delta(h^{jk}h_{jk})-4\lambda\delta (\Phi^2)\bigg).
\eea
The new term is
\bea
\tilde{S}^{(2)}_{add}=\frac{1}{16\pi G_N}\int [dxd\tilde{x}]\ \bigg(-\frac{\lambda}{4}b^{jk}b_{\mu\nu}-\frac{\lambda}{4}h^{jk}h_{jk}+4\lambda \Phi^2\bigg).
\eea
\\
The quadratic action for the massive states is: 
\bea
\tilde{S}^{(2)}_{DFT}  & =  &  \tilde{S}^{(2)}_{new} +  \tilde{S}^{(2)}_{add}  \nonumber \\ 
{} & =  & \tilde{S}^{(2)}_{new} + \frac{1}{16\pi G_N}\int [dxd\tilde{x}]\ \bigg(-\frac{\lambda}{4}b^{jk}b_{jk}-\frac{\lambda}{4}h^{jk}h_{jk}+4\lambda \Phi^2\bigg).
\eea
The gauge invariance requires the equality of the gauge parameters. 
It is consistent with the gauge transformation of the one-form gauge field. 
The constraint of the gauge parameters implies that the target space and its dual are not independent.  
\\

\noindent 
When the dilaton field vanishes, the scalar dilaton field at the quadratic order is 
\bea
\Phi=-\frac{h^j{}_j}{4}. 
\eea 
The familiar graviton mass term $\lambda\big(h^{jk}h_{jk}-(h^j{}_j)^2\big)$ appears in the quadratic theory. 
The non-vanishing $\lambda$ is due to $N_L\neq N_R$. 
It implies that the mass term is due to the simultaneous appearance of the momentum and winding modes. 
From the point of view of non-compact lower-dimensional spacetime, the squared mass of the graviton is 
\bea
{\cal M}_g^2=p^2+\omega^2+\frac{2(N_L-N_R)}{\alpha^{\prime}}. 
\eea 

\section{Cubic Theory}
\label{sec:5} 
\noindent 
We first show the gauge transformation and then derive the cubic action. 
The gauge invariance requires the coupling between the one-form gauge field and other fields. 
Although Double Field Theory does not have a consistent truncation, the cubic target space theory is self-consistent for each mode. 
Hence we truncate other modes and get the self-consistent theory even if the issue of truncation exists. 
We provide the details for constructing the cubic action in Appendix \ref{appb}.

\subsection{Gauge Transformation} 
\noindent
The gauge transformation with $\epsilon_j=\tilde{\epsilon}_j$ are \cite{Hull:2009mi}:
\bea
\delta e_{jk}=2\bar{D}_k\epsilon_j+
\big( (D_j\epsilon^l)e_{lk}- (D^l\epsilon_j)e_{lk}+\epsilon_lD^le_{jk}\big); \qquad 
\delta\Phi=-\frac{1}{2}D_j\epsilon^j+\epsilon_jD^j\Phi, 
\eea
where 
\bea
D_k\equiv\partial_k-\tilde{\partial}_k; \qquad \bar{D}_k\equiv\partial_k+\tilde{\partial}_k.
\eea
The $e_{jk}$ is a non-linear combination of $h_{jk}$ and $b_{jk}$ \cite{Hull:2009mi}
\bea
e_{jk}=h_{jk}+b_{jk}+\cdots. 
\eea 

\subsection{Cubic Action}
\noindent 
We also begin from the massless cubic action
\bea
&&S_{DFT}^{(2)}+S_{DFT}^{(3)}
\nn\\
&=&
\int \lbrack dxd\tilde{x}\rbrack\ 
\bigg\lbrack\frac{1}{4}e_{jk}\Box e^{jk}
+\frac{1}{4}(\bar{D}^ke_{jk})^2
+\frac{1}{4}(D^je_{jk})^2
-2\Phi D^j\bar{D}^ke_{jk}
-4\Phi\Box\Phi
\nn\\
&&
+\frac{1}{4}e_{jk}\bigg(
(D^j e_{lm})(\bar{D}^ke^{lm})
-(D^je_{lm})(\bar{D}^me^{lk})
-(D^l e^{jm})(\bar{D}^k e_{lm})\bigg)
\nn\\
&&
+\frac{1}{2}\Phi\bigg(
(D^j e_{jk})^2
+(\bar{D}^k e_{jk})^2
+\frac{1}{2}(D_l e_{jk})^2
+\frac{1}{2}(\bar{D}_le_{jk})^2
+2e^{jk}(D_jD^le_{lk}+\bar{D}_k\bar{D}^le_{jl})\bigg)
\nn\\
&&
+4e_{jk}\Phi D^j\bar{D}^k\Phi
+4\Phi^2\Box \Phi\bigg\rbrack,
\eea
where
\bea
\Box\equiv\frac{1}{2}(D^2+\bar{D}^2).
\eea 
We use the same corresponding of the massless and massive fields. 
Up to the cubic order, the non-invariant terms are:
\bea
&&
\delta \tilde{S}_{DFT}^{(2)}+
\delta \tilde{S}_{new}^{(3)}
\nn\\
&=&
\frac{\lambda}{16\pi G_N}\int \lbrack dxd\tilde{x}\rbrack\ 
\bigg\lbrack
-\frac{1}{4}\delta(\Phi e^{lk}e_{lk})
+4\delta(\Phi\Phi\Phi)
+\frac{1}{2}\epsilon^me_{mk}D_le^{lk}
+2\epsilon_le^{lk}\bar{D}_k\Phi
\bigg\rbrack
.
\nn\\
\eea 
For obtaining the gauge invariance, we need to add a new term 
\bea
&&
\tilde{S}_{add}^{(3)}
\nn\\
&=&
\frac{\lambda}{16\pi G_N}\int \lbrack dxd\tilde{x}\rbrack\ 
\bigg\lbrack
\frac{1}{4}\Phi e^{lk} e_{lk}
-4\Phi\Phi\Phi
\nn\\
&&
-\frac{1}{2}A^me_{mk}D_le^{lk}
-2A_le^{lk}\bar{D}_k\Phi
+A^m\bar{D}_k(A_m)D_l(e^{lk})
+2\lambda A_lA^l\Phi
\bigg\rbrack.
\eea
Hence we obtain the gauge symmetry for the massive state:
\bea
\delta\big( \tilde{S}_{DFT}^{(2)}+\tilde{S}_{new}^{(3)}+\tilde{S}_{add}^{(3)}\big)
\equiv
\delta\tilde{S}_{DFT}^{(2)}+\delta\tilde{S}_{DFT}^{(3)}=0.
\eea 
The cubic action becomes:
\bea
&&
\tilde{S}_{DFT}^{(2)}+\tilde{S}_{DFT}^{(3)}
\nn\\
&=&
\tilde{S}_{DFT}^{(2)}
+\tilde{S}_{new}^{(3)}
+\tilde{S}_{add}^{(3)}
\nn\\
&=&
\tilde{S}_{DFT}^{(2)}
+\tilde{S}_{new}^{(3)}
\nn\\
&&
+\frac{\lambda}{16\pi G_N}\int \lbrack dxd\tilde{x}\rbrack\ 
\bigg\lbrack
\frac{1}{4}\Phi e^{lk} e_{lk}
-4\Phi\Phi\Phi
\nn\\
&&
-\frac{1}{2}A^me_{mk}D_le^{lk}
-2A_le^{lk}\bar{D}_k\Phi
+A^m\bar{D}_k(A_m)D_l(e^{lk})
+2\lambda A_lA^l\Phi
\bigg\rbrack.
\eea
We write the details in Appendix \ref{appb}.

\section{$d=1$}
\label{sec:6}
\noindent 
The simplest solution should be for $d=1$. 
For one double direction, the constraints are: 
\bea
K_A^2=K_B^2=\lambda
\eea
give the momenta:
\bea
K_A=
\begin{pmatrix}
a
\\
\frac{\lambda}{2a}
\end{pmatrix}; \qquad 
K_B=
\begin{pmatrix}
b
\\
\frac{\lambda}{2b}
\end{pmatrix}, 
\eea 
where $a$ and $b$ are real-valued. 
The constraint for the product of the $K_A$ and $K_B$ is
\bea
K_AK_B=-\frac{\lambda}{2}
\eea
leads to the imaginary $a$ or $b$
\bea
\frac{a}{b}=\frac{-1\pm\sqrt{3}i}{2}.
\eea
Hence the contradiction shows no non-trivial solution for $d=1$. 
In other words, the only solution satisfying the strong constraint exists. 
For $\lambda=0$, ones can show the impossibility of the non-trivial solution \cite{Ma:2016vgq, Ma:2017jeq}. 
We generalize the proof to $\lambda\neq 0$.  
Our proof only requires the quadratic products of momenta. 
Therefore, it doe not suffer from any ambiguity. 
Hence only for $d>1$, it is possible to allow the massive states $\lambda\neq 0$. 
The result also shows the difference between String Theory and Double Field Theory. 
Double Field Theory for $d=1$ only has the modes of $N_L=N_R$. 
String Theory allows more states than Double Field Theory. 
The dual target spaces are relevant to the winding modes in Double Field Theory. 
Hence it gives more constraints to the choices of $N_L$ and $N_R$. 

\section{Discussion and Conclusion}
\label{sec:7} 
\noindent
We derived the cubic action for $N_L+N_R=2$ in Double Field Theory \cite{Hull:2009mi}. 
With $N_L\neq N_R$, the fields become massive. 
The field contents of the massive and massless cases have a one-to-one correspondence. 
Therefore, we could use the massless theory $(N_L=1, N_R=1)$ to build the massive case. 
From the perspective of String Theory, the constraints of fields should follow from the level matching condition. 
Therefore, the parameter \cite{Hull:2009mi}
\bea
\lambda=\frac{2(N_L-N_R)}{\alpha^{\prime}}
\eea 
generates the mass term and new interaction. 
We implemented the constraint to the triple products. 
Without being in contradiction with the integration by part, the generalization is unique. 
When one chooses $N_L\neq N_R$, the momentum and winding modes do not vanish simultaneously. 
The interaction of the momentum and winding modes generates a massive graviton. 
Although $N_L< N_R$ gives the vacuum instability, the cubic interaction has an opposite sign to avoid the issue. 
The stable vacuum is possible, similar to spontaneous symmetry breaking. 
The massive theory relies on a solution with $\lambda\neq 0$. 
We showed no such a solution for $d=1$. 
\\

\noindent
We obtained the massive gravity theory from $N_L\neq N_R$. 
The non-linear deformation of massive gravity theory suffered from the issue of a ghost mode. 
Our theory respects String Theory. 
Therefore, the target space theory should be self-consistent. 
Hence one can apply our study to cosmology observation. 
The graviton mass is relevant to the interaction of the winding and momentum modes. 
It is possible to show the experimental constraint to $\alpha^{\prime}$.  
\\ 

\noindent 
Up to the fluctuation level, it is hard to understand the perspective of geometry. 
We could realize the gauge symmetry up to the cubic order. 
Therefore, it is possible to extend our study to the generalized metric formulation \cite{Hohm:2010pp}. 
For the massless theory, the generalized metric plays the role of manifest T-duality. 
The massive state should not have the O($D$, $D$) symmetric structure. 
However, one can always choose symmetric and anti-symmetric parts to form the generalized metric. 
It provides the convenient O($D$, $D$) indices to write action without changing the gauge transformation. 
One can explore the geometry by performing a similar analysis to the generalized metric formulation.   
\\

\noindent 
In the end, we comment on the consistent truncation. 
We first choose our consideration $N_L+N_R=2$. 
The squared of mass is
\bea
{\cal M}^2=\frac{n^2}{R^2}+m^2\frac{R^2}{\alpha^{\prime 2}}.
\eea 
We then compare the squared of mass to the $N_L+N_R=3$ case. 
One additional $1/\alpha^{\prime}$ term appears in the mass scale
\bea
{\cal M}^2=\frac{n^2}{R^2}+m^2\frac{R^2}{\alpha^{\prime 2}}+\frac{2}{\alpha^{\prime}}.
\eea 
The choice $(n, m)=(2, 1)$ in $N_L+N_R=2$ has a higher mass scale than $(n, m)=(1, 3)$ in $N_L+N_R=3$ when the compactified radius is small enough
\bea
R^2\le\frac{\alpha^{\prime}}{2}.
\eea 
Hence concerning the $N_L+N_R=2$ is not enough for the consistent truncation (energy). 
The summation of infinite modes is necessary.  
This issue suggests that each mode in Double Field Theory cannot form a self-consistent theory. 
Since the quadratic-order corresponds to the non-interacting theory, one should expect gauge invariance for each mode. 
For the non-linear theory, it is hard to expect gauge invariance.  
Hence we extend our study to the cubic level. 
We can introduce the new coupling to show a gauge-invariant theory. 
Our results showed a contradiction to the expectation. 
Each mode can form a self-consistent target space theory even with no consistent truncation in String Theory. 
 
\section*{Acknowledgments}
\noindent
The author thanks David S. Berman, Chi-Ming Chang, Xing Huang, Jeong-Hyuck Park, and Franco Pezzella for their discussion. 
The author also would like to thank for Nan-Peng Ma for his encouragement.
\\

\noindent
Chen-Te Ma acknowledges the China Postdoctoral Science Foundation, Postdoctoral General Funding: Second Class (Grant No. 2019M652926); 
Foreign Young Talents Program (Grant No. QN20200230017); 
Post-Doctoral International Exchange Program.

\appendix
\section{Closed Gauge Transformation of\\
 Generalized Metric}
\label{appa}
\noindent
We provide the details to the gauge transformation of the generalized metric: 
\bea
&&(\delta_{\xi_1}*\delta_{\xi_2}){\cal H}^{MN}
\nn\\
&=&\xi_2^P*\partial_P\xi_1^Q*\partial_Q{\cal H}^{MN}+\xi_2^P*\xi_1^Q*\partial_P\partial_Q{\cal H}^{MN}
\nn\\
&&+\xi_2^P*\bigg(\partial_P\partial^M\xi_{1Q}-\partial_P\partial_Q\xi_1^M\bigg)*{\cal H}^{QN}
+\xi_2^P*\bigg(\partial^M\xi_{1Q}-\partial_Q\xi_1^M\bigg)*\partial_P{\cal H}^{QN}
\nn\\
&&+\xi_2^P*\bigg(\partial_P\partial^N\xi_{1Q}-\partial_P\partial_Q\xi_1^N\bigg)*{\cal H}^{MQ}
+\xi_2^P*\bigg(\partial^N\xi_{1Q}-\partial_Q\xi_1^N\bigg)*\partial_P{\cal H}^{MQ}
\nn\\
&&+\bigg(\partial^M\xi_{2P}-\partial_P\xi_2^M\bigg)*\xi_1^Q*\partial_Q{\cal H}^{PN}
\nn\\
&&+\bigg(\partial^M\xi_{2P}-\partial_P\xi_2^M\bigg)*\bigg(\partial^P\xi_{1Q}-\partial_Q\xi_1^P\bigg)*{\cal H}^{QN}
\nn\\
&&+\bigg(\partial^M\xi_{2P}-\partial_P\xi_2^M\bigg)*\bigg(\partial^N\xi_{1Q}-\partial_Q\xi_1^N\bigg)*{\cal H}^{PQ}
\nn\\
&&+\bigg(\partial^N\xi_{2P}-\partial_P\xi_2^N\bigg)*\xi_1^Q*\partial_Q{\cal H}^{PM}
\nn\\
&&+\bigg(\partial^N\xi_{2P}-\partial_P\xi_2^N\bigg)*\bigg(\partial^P\xi_{1Q}-\partial_Q\xi_1^P\bigg)*{\cal H}^{QM}
\nn\\
&&+\bigg(\partial^N\xi_{2P}-\partial_P\xi_2^N\bigg)*\bigg(\partial^M\xi_{1Q}-\partial_Q\xi_1^M\bigg)*{\cal H}^{PQ},
\eea
\bea
&&\lbrack\delta_{\xi_1}, \delta_{\xi_2}\rbrack_*{\cal H}^{MN}
\nn\\
&=&\xi_2^P*\partial_P\xi_1^Q*\partial_Q{\cal H}^{MN}
\nn\\
&&+\xi_2^P*\bigg(\partial_P\partial^M\xi_{1Q}-\partial_P\partial_Q\xi_1^M\bigg)*{\cal H}^{QN}
\nn\\
&&+\xi_2^P*\bigg(\partial_P\partial^N\xi_{1Q}-\partial_P\partial_Q\xi_1^N\bigg)*{\cal H}^{MQ}
\nn\\
&&+\bigg(\partial^M\xi_{2P}-\partial_P\xi_2^M\bigg)*\bigg(\partial^P\xi_{1Q}-\partial_Q\xi_1^P\bigg)*{\cal H}^{QN}
\nn\\
&&+\bigg(\partial^N\xi_{2P}-\partial_P\xi_2^N\bigg)*\bigg(\partial^P\xi_{1Q}-\partial_Q\xi_1^P\bigg)*{\cal H}^{QM}
\nn\\
&&-\xi_1^P*\partial_P\xi_2^Q*\partial_Q{\cal H}^{MN}
\nn\\
&&-\xi_1^P*\bigg(\partial_P\partial^M\xi_{2Q}-\partial_P\partial_Q\xi_2^M\bigg)*{\cal H}^{QN}
-\xi_1^P*\bigg(\partial_P\partial^N\xi_{2Q}-\partial_P\partial_Q\xi_2^N\bigg)*{\cal H}^{MQ}
\nn\\
&&-\bigg(\partial^M\xi_{1P}-\partial_P\xi_1^M\bigg)*\bigg(\partial^P\xi_{2Q}-\partial_Q\xi_2^P\bigg)*{\cal H}^{QN}
\nn\\
&&-\bigg(\partial^N\xi_{1P}-\partial_P\xi_1^N\bigg)*\bigg(\partial^P\xi_{2Q}-\partial_Q\xi_2^P\bigg)*{\cal H}^{QM}
\nn\\
&=&-\lbrack\xi_1, \xi_2\rbrack_C^P*\partial_P{\cal H}^{MN}-\bigg(\partial^M\lbrack\xi_1, \xi_2\rbrack_{C P}-\partial_P\lbrack\xi_1, \xi_2\rbrack_C^M\bigg)*{\cal H}^{PN}
\nn\\
&&-\bigg(\partial^N\lbrack\xi_1, \xi_2\rbrack_{C P}-\partial_P\lbrack\xi_1, \xi_2\rbrack_C^N\bigg)*{\cal H}^{MP}
\nn\\
&=&-\delta_{\lbrack\xi_1, \xi_2\rbrack_C}{\cal H}^{MN}.
\eea
Therefore, the gauge transformation is closed. 
In the calculation, we used the following results:
\bea
&&-\lbrack\xi_1, \xi_2\rbrack_C^P\partial_P{\cal H}^{MN}
\nn\\
&=&-\bigg(\xi_1^Q*\partial_Q\xi_2^P-\xi_2^Q*\partial_Q\xi_1^P
-\frac{1}{2}\eta^{PR}\eta_{OQ}\xi_1^O*\partial_R\xi_2^Q+\frac{1}{2}\eta^{PR}\eta_{OQ}\xi_2^O*\partial_R\xi_1^Q\bigg)*\partial_P{\cal H}^{MN}
\nn\\
&=&\bigg(\xi_2^Q*\partial_Q\xi_1^P-\xi_1^Q*\partial_Q\xi_2^P\bigg)*\partial_P{\cal H}^{MN},
\eea
\bea
&&-\bigg(\partial^M\lbrack\xi_1, \xi_2\rbrack_{C P}-\partial_P\lbrack\xi_1, \xi_2\rbrack_C^M\bigg)*{\cal H}^{PN}
\nn\\
&=&-\partial^M\bigg(\xi_1^Q*\partial_Q\xi_{2P}-\xi_2^Q*\partial_Q\xi_{1P}-\frac{1}{2}\eta_{OQ}\xi_1^O*\partial_P\xi_2^Q+\frac{1}{2}\eta_{OQ}\xi_2^O*\partial_P\xi_1^Q\bigg)*{\cal H}^{PN}
\nn\\
&&+\partial_P\bigg(\xi_1^Q*\partial_Q\xi_{2}^M-\xi_2^Q*\partial_Q\xi_{1}^M-\frac{1}{2}\eta_{OQ}\xi_1^O*\partial^M\xi_2^Q+\frac{1}{2}\eta_{OQ}\xi_2^O*\partial^M\xi_1^Q\bigg)*{\cal H}^{PN}
\nn\\
&=&\bigg(-\partial^M\xi_1^Q*\partial_Q\xi_{2P}-\xi^Q_1*\partial^M\partial_Q\xi_{2P}
+\partial^M\xi_2^Q*\partial_Q\xi_{1P}+\xi_2^Q*\partial^M\partial_Q\xi_{1P}
\nn\\
&&+\frac{1}{2}\partial^M\xi_{1Q}*\partial_P\xi_2^Q+\frac{1}{2}\xi_{1Q}*\partial^M\partial_P\xi_2^Q
-\frac{1}{2}\partial^M\xi_{2Q}*\partial_P\xi_1^Q-\xi_{2Q}*\partial^M\partial_P\xi_1^Q\bigg)*{\cal H}^{PN}
\nn\\
&&+\bigg(\partial_P\xi_1^Q*\partial_Q\xi_{2}^M+\xi^Q_1*\partial_P\partial_Q\xi_{2}^M
-\partial_P\xi_2^Q*\partial_Q\xi_{1}^M-\xi_2^Q*\partial_P\partial_Q\xi_{1}^M
\nn\\
&&-\frac{1}{2}\partial_P\xi_{1Q}*\partial^M\xi_2^Q-\frac{1}{2}\xi_{1Q}*\partial_P\partial^M\xi_2^Q
+\frac{1}{2}\partial_P\xi_{2Q}*\partial^M\xi_1^Q+\xi_{2Q}*\partial_P\partial^M\xi_1^Q\bigg)*{\cal H}^{PN},
\nn\\
\eea
and 
\bea
&&-\bigg(\partial^N\lbrack\xi_1, \xi_2\rbrack_{C P}-\partial_P\lbrack\xi_1, \xi_2\rbrack_C^N\bigg)*{\cal H}^{PM}
\nn\\
&=&\bigg(-\partial^N\xi_1^Q*\partial_Q\xi_{2P}-\xi^Q_1*\partial^N\partial_Q\xi_{2P}
+\partial^N\xi_2^Q*\partial_Q\xi_{1P}+\xi_2^Q*\partial^N\partial_Q\xi_{1P}
\nn\\
&&+\frac{1}{2}\partial^N\xi_{1Q}*\partial_P\xi_2^Q+\frac{1}{2}\xi_{1Q}*\partial^N\partial_P\xi_2^Q
-\frac{1}{2}\partial^N\xi_{2Q}*\partial_P\xi_1^Q-\xi_{2Q}*\partial^N\partial_P\xi_1^Q\bigg)*{\cal H}^{PM}
\nn\\
&&+\bigg(\partial_P\xi_1^Q*\partial_Q\xi_{2}^N+\xi^Q_1*\partial_P\partial_Q\xi_{2}^N
-\partial_P\xi_2^Q*\partial_Q\xi_{1}^N-\xi_2^Q*\partial_P\partial_Q\xi_{1}^N
\nn\\
&&-\frac{1}{2}\partial_P\xi_{1Q}*\partial^N\xi_2^Q-\frac{1}{2}\xi_{1Q}*\partial_P\partial^N\xi_2^Q
+\frac{1}{2}\partial_P\xi_{2Q}*\partial^N\xi_1^Q+\xi_{2Q}*\partial_P\partial^N\xi_1^Q\bigg)*{\cal H}^{PM},
\nn\\
\eea
which is obtained from 
\bea
-\bigg(\partial^M\lbrack\xi_1, \xi_2\rbrack_{C P}-\partial_P\lbrack\xi_1, \xi_2\rbrack_C^M\bigg)*{\cal H}^{PN}
\eea
by exchanging the index $M$ and the index $N$. 

\section{Cubic Action for Massive State}
\label{appb}
\noindent
The gauge transformation of the quadratic fields provides the following terms:
\bea
&&
\delta \tilde{S}_{DFT}^{(2)}
\nn\\
&=&
\frac{\lambda}{16\pi G_N}\int [dxd\tilde{x}]\ \bigg(
\frac{1}{4}\epsilon^m e_{mk}D_le^{lk}
-\Phi e^{lk} \bar{D}_k\epsilon_l
-\Phi\epsilon_l\bar{D}_ke^{lk}
\nn\\
&&
-\frac{1}{2}e^{jk}\big(e_{lk} D_j\epsilon^l-e_{lk} D^l\epsilon_j+\epsilon_l D^le_{jk}\big)
+8\epsilon_j\Phi D^j\Phi
\bigg)
\nn\\
&=&\frac{\lambda}{16\pi G_N}\int [dxd\tilde{x}]\ \bigg(
-\frac{1}{2} e^{lk} e_{lk}\delta\Phi
+\frac{8}{3}\delta(\Phi\Phi\Phi)
+\frac{1}{4}\epsilon^m e_{mk}D_le^{lk}
+\epsilon_l e^{lk}\bar{D}_k\Phi
\bigg).
\nn\\
\eea
For the cubic fields, the gauge transformation provides the following non-invariant terms:
\bea
&&
\delta \tilde{S}_{new}^{(3)}
\nn\\
&=&
\frac{1}{16\pi G_N}\int \lbrack dxd\tilde{x}\rbrack\ 
\bigg\lbrack 
\frac{1}{2}(\bar{D}_k\epsilon_j)(D^je_{lm})(\bar{D}^ke^{lm})
-\frac{1}{2}e_{jk}(D^j\bar{D}_m\epsilon_l)(\bar{D}^me^{lk})
\nn\\
&&
-\frac{1}{2}(\bar{D}_k\epsilon_j)(D^le^{jm})(\bar{D}^ke_{lm})
\nn\\
&&
+\Phi\bigg(2(\bar{D}^k\bar{D}_k\epsilon_j)(\bar{D}_le^{jl})
+2(\bar{D}^k\epsilon^j)(\bar{D}_k\bar{D}^le_{jl})
+2e^{jk}\bar{D}_k\bar{D}^l\bar{D}_l\epsilon_j\bigg)
+8(\bar{D}_k\epsilon_j)\Phi D^j\bar{D}^k\Phi
\bigg\rbrack
\nn\\
&=&
\frac{\lambda}{16\pi G_N}\int \lbrack dxd\tilde{x}\rbrack\ 
\bigg\lbrack 
\frac{1}{4}\epsilon_je^{lm}D^je_{lm}
-\frac{1}{4}e_{jk}e^{lk}D^j\epsilon_l
-\frac{1}{4}\epsilon_je_{lm}D^le^{jm}
\nn\\
&&
-2\epsilon_j\Phi\bar{D}_le^{jl}
+\epsilon^j\Phi\bar{D}^le_{jl}
-2\Phi e^{jk}\bar{D}_k\epsilon_j
+4\epsilon_j\Phi D^j\Phi
\bigg\rbrack
\nn\\
&=&
\frac{\lambda}{16\pi G_N}\int \lbrack dxd\tilde{x}\rbrack\ 
\bigg\lbrack 
\frac{1}{4}e^{lm}e_{lm}\delta\Phi
-\frac{1}{2}\Phi\delta(e^{jk}e_{jk})
+\frac{1}{4}\epsilon_je^{jm}D^le_{lm}
\nn\\
&&
-\epsilon_j\Phi\bar{D}_le^{jl}
+\frac{4}{3}\delta(\Phi\Phi\Phi)
\bigg\rbrack.
\eea 
Hence up to the cubic fields, the non-invariant terms are:
\bea
&&
\delta \tilde{S}_{DFT}^{(2)}+
\delta \tilde{S}_{new}^{(3)}
\nn\\
&=&
\frac{\lambda}{16\pi G_N}\int \lbrack dxd\tilde{x}\rbrack\ 
\bigg\lbrack
-\frac{1}{2} e^{lk} e_{lk}\delta\Phi
+\frac{8}{3}\delta(\Phi\Phi\Phi)
+\frac{1}{4}\epsilon^m e_{mk}D_le^{lk}
+\epsilon_l e^{lk}\bar{D}_k\Phi
\nn\\
&&
+\frac{1}{4}e^{lm}e_{lm}\delta\Phi
-\frac{1}{2}\Phi\delta(e^{jk}e_{jk})
+\frac{1}{4}\epsilon_je^{jm}D^le_{lm}
-\epsilon_j\Phi\bar{D}_le^{jl}
+\frac{4}{3}\delta(\Phi\Phi\Phi)
\bigg\rbrack
\nn\\
&=&
\frac{\lambda}{16\pi G_N}\int \lbrack dxd\tilde{x}\rbrack\ 
\bigg\lbrack
-\frac{1}{4}\delta(\Phi e^{lk}e_{lk})
+4\delta(\Phi\Phi\Phi)
+\frac{1}{2}\epsilon^me_{mk}D_le^{lk}
+2\epsilon_le^{lk}\bar{D}_k\Phi
\bigg\rbrack
.
\nn\\
\eea 
\\

\noindent
For the gauge invariance, we need to add a new term 
\bea
&&
\tilde{S}_{add}^{(3)}
\nn\\
&=&
\frac{\lambda}{16\pi G_N}\int \lbrack dxd\tilde{x}\rbrack\ 
\bigg\lbrack
\frac{1}{4}\Phi e^{lk} e_{lk}
-4\Phi\Phi\Phi
\nn\\
&&
-\frac{1}{2}A^me_{mk}D_le^{lk}
-2A_le^{lk}\bar{D}_k\Phi
+A^m\bar{D}_k(A_m)D_l(e^{lk})
+2\lambda A_lA^l\Phi
\bigg\rbrack.
\eea
The gauge transformation acting on the new term shows:
\bea
&&
\delta\tilde{S}_{add}^{(3)}
\nn\\
&=&
\frac{\lambda}{16\pi G_N}\int[dxd\tilde{x}]\ \bigg\lbrack
\frac{1}{4}\delta(\Phi e^{lk}e_{lk})
-4\delta(\Phi\Phi\Phi)
-\frac{1}{2}\epsilon^me_{mk}D_l(e^{lk})
-2\epsilon_le^{lk}\bar{D}_k\Phi
\nn\\
&&
-\frac{1}{2}\delta\big(\bar{D}_k(A^mA_m)\big)D_le^{lk}
-2\lambda\delta(A_lA^l)\Phi
\nn\\
&&
+\frac{1}{2}\delta\big(\bar{D}_k(A^mA_m)\big)D_le^{lk}
+\lambda A^mA_mD_l\epsilon^l
\nn\\
&&
+2\lambda\delta(A_lA^l)\Phi
-\lambda A_lA^lD_j\epsilon^j
\bigg\rbrack
\nn\\
&=&
\frac{\lambda}{16\pi G_N}\int[dxd\tilde{x}]\ \bigg\lbrack
\frac{1}{4}\delta(\Phi e^{lk}e_{lk})
-4\delta(\Phi\Phi\Phi)
-\frac{1}{2}\epsilon^me_{mk}D_l(e^{lk})
-2\epsilon_le^{lk}\bar{D}_k\Phi
\bigg\rbrack,
\nn\\
\eea 
Hence we show the gauge symmetry for the cubic action:
\bea
\delta\big( \tilde{S}_{DFT}^{(2)}+\tilde{S}_{new}^{(3)}+\tilde{S}_{add}^{(3)}\big)
\equiv
\delta\tilde{S}_{DFT}^{(2)}+
\delta\tilde{S}_{DFT}^{(3)}=0.
\eea 
\\

\noindent
The cubic action for the massive state is:
\bea
&&
\tilde{S}_{DFT}^{(2)}
+\tilde{S}_{DFT}^{(3)}
\nn\\
&=&
\tilde{S}_{DFT}^{(2)}
+\tilde{S}_{new}^{(3)}
+\tilde{S}_{add}^{(3)}
\nn\\
&=&
\tilde{S}_{DFT}^{(2)}
+\tilde{S}_{new}^{(3)}
\nn\\
&&
+\frac{\lambda}{16\pi G_N}\int \lbrack dxd\tilde{x}\rbrack\ 
\bigg\lbrack
\frac{1}{4}\Phi e^{lk} e_{lk}
-4\Phi\Phi\Phi
\nn\\
&&
-\frac{1}{2}A^me_{mk}D_le^{lk}
-2A_le^{lk}\bar{D}_k\Phi
+A^m\bar{D}_k(A_m)D_l(e^{lk})
+2\lambda A_lA^l\Phi
\bigg\rbrack.
\eea

\end{document}